\documentclass[10pt,technote,onecolumn]{article} \usepackage{graphicx}

\usepackage{amsmath,amssymb,amsfonts} 

\newtheorem{theorem}{Theorem}[section]  
\newtheorem{lemma}[theorem]{Lemma}

\title{Best Response Games on Regular Graphs} \author{Richard Southwell and Chris Cannings} \date{January 2013} \begin{document}
   \maketitle

\begin{abstract} 

With the growth of the internet it is becoming increasingly important to understand how the behaviour of players is affected by the topology of the network
interconnecting them. Many models which involve networks of interacting players have been proposed and best response games are amongst the simplest. In best
response games each vertex simultaneously updates to employ the best response to their current surroundings. We concentrate upon trying to understand the
dynamics of best response games on regular graphs with many strategies. When more than two strategies are present highly complex dynamics can ensue. We focus
upon trying to understand exactly how best response games on regular graphs sample from the space of possible cellular automata. To understand this issue we
investigate convex divisions in high dimensional space and we prove that almost every division of $k-1$ dimensional space into $k$ convex regions includes a
single point where all regions meet. We then find connections between the convex geometry of best response games and the theory of alternating circuits on
graphs. Exploiting these unexpected connections allows us to gain an interesting answer to our question of when cellular automata are best response games.

\end{abstract} \section{Introduction}

Game theory is a rich subject which has become more diverse over the years. Early game theory \cite{von} focused on how rational players would behave in
strategic conflicts. Concepts like the Nash equilibrium \cite{nash} helped theorists to predict final outcomes of games, and this benefited areas like economics.
A more recent development is evolutionary game theory \cite{may}. Rather than assuming that players are hyper-rational, evolutionary game theory concerns itself
with large populations of players that change their strategies via simple selection mechanisms. Players repeatedly engage in games with other members of the
population and the way the populations strategies evolve depends upon the selection mechanism employed.

One way to introduce a spatial aspect into evolutionary games is to imagine the players as vertices within a graph \cite{gog}, \cite{sant}.
 The links represent interactions so vertices play games with their neighbours. The players adapt their strategies over time to try to increase their success
 against neighbours.
Many kinds of update rules have been investigated. These include imitation, where vertices imitate their most successful neighbour, and best response, where
vertices update to employ the strategy best suiting their current surroundings.
 In \cite{nov} the authors study a two dimensional cellular automata with update rules based upon games, beautiful patterns emerge.

The update mechanism that we concentrate upon is best response, where vertices update to employ the strategy that maximisie their total payoff in a game with
each neighbour. The vertices update their strategies myopically and synchronously. The strategy a vertex updates to may not be optimal because neighbouring
vertices change their strategies at the same time.
 Work on these kind of systems includes \cite{house1} and \cite{house2}, where it was shown that a two strategy game running upon any graph will eventually reach

 a fixed point or period two orbit. Two strategy best response games on various graph structures were also studied in \cite{newcan}.

The systems we consider are essentially cellular automata with update rules that are induced by the details of the game. In our consideration of best response
games on the circle we will see many of Wolfram's 256 elementary cellular automata \cite{nks} appear. Under these models the states of the vertices on a circle
or line graph take values in $\{ 0 , 1 \}$. The states of vertices change with time so that the future state of any cell depends upon the current state of itself
and its neigbours. Each systems is specified by a mapping $f:\{ 0 , 1 \} ^3 \mapsto \{ 0 , 1 \}$ so that $f( x_{i-1} , x_i , x_{i+1})$ is the future state of a
vertex in state $x_i$ with neighbours in states $x_{i-1}$ and $x_{i+1}$ to its left and right. Each system is indexed with a number $f(0,0,0).2^0 + f(0,0,1).2^1
+ f(0,1,0).2^2 + f(0,1,1).2^3 + f(1,0,0).2^4 + f(1,0,1).2^5 + f(1,1,0).2^6 + f(1,1,1).2^7$.

We concentrate upon games on the circle because they provide the easiest ways to illustrate our results. Our methods can easily be extended to other graph
structures, even non-regular ones.

\subsection{Definitions}

For a set $S$ let $\mathbb{P}^d (S)$ to be the set of size $d$ multi sets of elements from $S$ i.e. $\mathbb{P}^d (S)$ is the set of all unordered $d$-tuples
$\{s_1 , s_2 , ...., s_d \}$ of members of $S$, including those $d$-tuples containing more than one of the same element. Let us define a regular automata as a
quad $(G,S, \Phi ^0 (G),F)$, where $G=(V,E)$ is a regular degree $d$ graph, $S$ is a set of states, $\Phi ^0(G)$ is an assignment of a state $\phi ^0 (v) \in S$
to each vertex $v \in V$ (the initial configuration) and $F: \mathbb{P}^d (S) \mapsto S$ is the update function/rule.

Such a regular automata evolves so that at time step $t \in \mathbb{N} _0$, the future state of a vertex $v$, at time $t+1$, will be \begin{equation} \phi^{t+1}
(v) = F ( \{ s_1,s_2,...,s_d \} ) \end{equation} where \begin{equation} \{ s_1,s_2,...,s_d \}= \{ \phi^t (u) | u \in Ne(v) \} \end{equation} is the set of states
of vertices in $v$'s neighbourhood, $Ne(v)$, at time $t$.

A game $( \zeta , { \bf M })$ consists of a set of strategies $\zeta=\{1,2,...,k\}$ which we label with integers, together with a $k \times k$ payoff matrix ${
\bf M }$ such that ${ \bf M}_{i,j} \in \mathbb{R}$ is the payoff that a player receives from employing the $i$th strategy against the $j$th strategy.

The best response games we consider take place on a degree $d$ regular graph $G=(V,E)$. At time $t$ each vertex $v \in V$ employs a strategy $\phi ^t (v) \in
\zeta$. The total payoff \begin{equation} { \displaystyle \sum_{u \in Ne(v)} { \bf M}_{\phi^t (v) , \phi^t (u)}} \end{equation} of $v$ at time $t$ is the sum of
payoffs that $v$ receives from using its strategy in a game with each neighboring player. At time $t$ each vertex simultaneously updates its strategy, so that at
time $t+1$ it will employ the strategy that would have maximized its total payoff, given what its neighbours played at time $t$. In other words each player
updates to play the best response to its current surroundings.

Best response games on regular graphs are quads $(G,\zeta,\Phi ^0(G),{ \bf M})$,  which are defined by a regular graph $G$, a game $( \zeta , { \bf M })$ and an
assignment $\Phi ^0 (G)$ of an initial strategy to $\phi^0 (v) \in \zeta$ to each vertex of $G$. A best response game is a regular automata  $(G,\zeta
,\Phi^0(G),F)$ with an update function $F$ such that, for every unordered $d$-tuple of strategies $\{ s_1,s_2,...,s_d \} \in \mathbb{P} ^d ( \zeta )$, we have
$F( \{ s_1,s_2,...,s_d \} ) $ is the strategy in $ \{ \alpha \in \zeta \}$ that maximizes \begin{equation} { \displaystyle \sum _{i=1} ^d { \bf M } _ {\alpha ,
s_i } }. \end{equation} We refer to $F$ as the update function \emph{induced} by the game $(\zeta, { \bf M})$.
 Note that in rare cases the payoff matrix could be such that two strategies are tied as best responses to a possible local strategy configuration $\{s_1 , s_2
 ,...,s_d \}$. In such a case $F( \{ s_1,s_2,...,s_d \} )$ is not properly defined. Such a problem can always be alleviated by an infinitesimal perturbation of
 the elements in the payoff matrix ${ \bf M }$. We always assume our matrix is such that this problem never occurs.

\subsection{Examples on the circle}

The circle graph $C_n$ has vertices $\{ v_0 , v_1 ,...,v_{n-1} \}$ with $v_i$ is adjacent to $v_j$ if and only if $|i-j|=1$ $\mod {n}$. Every best response game
$(C_n , \zeta, \Phi ^0 (C_n) , { \bf M } )$ on the circle is a regular automata $(C_n , \zeta, \Phi ^0 (C_n) , F)$, where the strategy $\phi ^{t+1} (v_i)$
employed by vertex $v_i$ at time $t+1$ will be $F( \{ \phi ^t (v_{(i-1 \mod{n}) } ) , \phi ^t (v_{(i+1 \mod{n}) } \} )$, which is the strategy that maximizes the
vertices's total payoff against its neighbours strategies $\phi ^t (v_{(i-1 \mod{n}) } )$ and $\phi ^t (v_{(i+1 \mod{n}) } )$.

\begin{figure}
 \centering
\includegraphics[scale=0.4]{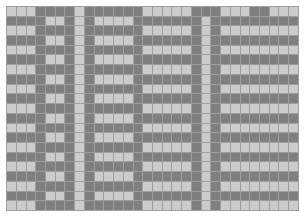} \caption{A space time plot showing the evolution of the Hawk - Dove example game on the circle.}
\label{hawkdovespace} \end{figure}

The payoff matrix

\begin{equation} {\bf M}=
 \left(
  \begin{array}{cc}
    1 & 0  \\
    4 & -2  \\
  \end{array}
\right) \label{hawkeqn} \end{equation}

 is a Hawk - Dove game, where $1$ is the passive dove strategy and $2$ is the aggressive hawk strategy.
Let us consider the best response game $(C_n , \zeta , \Phi ^0 (C_n) , { \bf M })$, where $\zeta =\{1,2\}$ and ${ \bf M }$ is as above. This corresponds to the
regular automata $(C_n , \zeta , \Phi ^0 (C_n) , F)$ with update function $F ( \{1,1 \} )= 2$, $F( \{1,2 \} ) = 2$ and $F ( \{2,2 \}) = 1$. We picture how this
automata will evolves on a 30 vertex circle from a random strategy configuration, with the space time plot pictured in figure \ref{hawkdovespace}.

A space time plot \cite{nks} is a grid where the $x$ axis is the index of the vertex of the circle and the $y$ axis (reading downwards) is the time step. Light
gray blocks denote employers of strategy 1 whilst dark blocks represent employers of strategy 2. In our game a player adjacent to an employer of dove and
employer of hawk gets updated to play the hawk strategy, hence if a block on our space time plot has a dark gray block on its left and a light gray block on its
right then the block below it will be dark gray. The system corresponds to Wolfram's cellular automata number 95, which is well known to have simple dynamics.
Every initial configuration evolving quickly to a fixed point or period 2 orbit. This system is an example of a threshold game \cite{house1}, \cite{house2}.

More complicated dynamics occur under the 3 strategy game with payoff matrix

\begin{equation} {\bf M}=
 \left(
  \begin{array}{ccc}
    3 & 94 & 46 \\
    33 & 85 & 66 \\
    52 & 2 & 67 \\
  \end{array}
\right), \label{foresteqn} \end{equation}

\begin{figure}
 \centering
\includegraphics[scale=0.6]{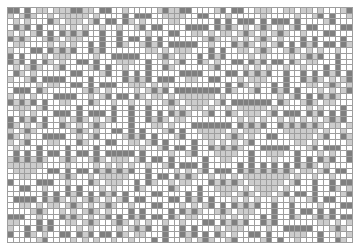} \caption{A space time plot showing the evolution of the three strategy game described by the payoff matrix in
equation \ref{foresteqn}.} \label{forestrand} \end{figure}

Figure \ref{forestrand} shows the evolution of the system from a random initial configuration on a 60 vertex circle over 40 time steps. In this case dark gray,
white and light gray blocks represent strategies 1, 2 and 3 respectively. The resulting cellular automata can be reduced to Wolfram's cellular automata number 90
which has been proven to be chaotic \cite{kurka} when running on an infinite circle.

In this paper we will enumerate all the update functions $F$ that can be induced by 2,3 or 4 strategy best response games on the circle by consideration of the
convex geometry behind best response game. Understanding this convex geometry allows one to understand how the update rules of arbitrary best response games are
induced.

\ \\

\section{Convex geometry behind best response games}\ \\ \label{sec:convgeo} \ \\

For a game $(\zeta , { \bf M } )$, let us think of each strategy $i \in \zeta = \{1,2,...,k\}$ as a unit vector \begin{equation} { \bf  e }(i)   = ( \delta
_{1,i} , \delta _ {2,i} , ...., \delta _{k,i}) \in \mathbb{R} ^k, \end{equation} where $\delta _{i,j}$ is the Kronecker delta.

The strategy space $\Delta \subseteq \mathbb{R} ^ k $ is the convex hull of the set of unit strategy vectors. The points \begin{equation} { \bf x } = (x_1 , x_2
,...,x_k) \in \Delta \end{equation} are probability distributions over the set of strategies $\zeta$ so that $x_i$ is the fraction of the $i$th strategy employed
in the strategy vector ${ \bf x }$.

The affine hull of $\{ { \bf e} (i)  | i \in \zeta \}$ forms the extended strategy space ${ \bf S}$ which we can write as \begin{equation} { \bf  S } = \{ { \bf
x } \in \mathbb{R} ^k | { \bf x 1 } ^T = { \bf 1 } \}, \end{equation} where ${ \bf 1 }$ is the length $k$ vector with each entry equal to $1$. Here ${ \bf S }$
is isomorphic to $\mathbb{R} ^{k-1}$ and $\Delta \subset { \bf S }$ so we think of $\Delta$ as a $k-1$ dimensional unit simplex, with the $k$ unit strategy
vectors ${ \bf e } (i) $ as its vertices. For each pair of strategy vectors $ \textbf{x} , \textbf{y} \in { \bf S }$ the payoff one receives from playing
$\textbf{x}$ against $\textbf{y}$ is \begin{equation} \textbf{x} { \bf M } \textbf{y} ^T. \end{equation}

The best response set to any strategy vector $\textbf{x} \in { \bf S }$ is the set of pure strategies $i$ such that $\forall j \in \zeta$ \begin{equation} { \bf
e }(i)  { \bf M }  \textbf{y} ^T \geq { \bf  e }(j) { \bf M  } \textbf{y} ^T. \end{equation}

The $i$th best response region ${ \bf R } _i  \subseteq { \bf S }$ is the set of all points in ${ \bf S }$ that have a best response set equal to $i \in \zeta$,
in other words ${ \bf R} _i $ is the set of points ${\bf x} \in {\bf S}$ where \begin{equation} [{\bf M}{\bf x}]_i \geq [{\bf M}{\bf x}]_j, \forall j \in \zeta -
i. \end{equation}

Let us define a `division' of a subset of Euclidian space to be a collection of closed regions such that every point of the space lies within some region, and the interiors of any pair of distinct regions do not intersect. We say a division is convex when each of its regions is convex. Every $k$ strategy game $( \zeta , { \bf M })$ induces a division of ${ \bf S }$ into $m \leq k$ convex best response regions ${ \bf R } _i $, because every point of ${ \bf S }$ belongs to some best response region ${ \bf R } _i $, and pairs of distinct best response regions only intersect at their boundaries.

Let ${ \bf T} _d  \subset \Delta$ be the set of all points ${ \bf x } \in \Delta$ that can be written as \begin{equation} \frac{ { \displaystyle \sum_{i \in D} }
{ \bf e } (i)  } {d} \end{equation} for some $D \in \mathbb{P} ^d ( \zeta )$ (where our sum takes into account that some elements may occur in $D$ several
times). The set of points ${ \bf T} _d $ will be partitioned into different best response regions ${ \bf R} _i  \cap { \bf T} _d $ and the nature of this
partition determines the update function $F$ of the regular automata that occurs when the game $(\zeta , { \bf M })$ is used for a best response game on a $d$
regular graph. We say that the partition of ${ \bf T} _d $ into different best response regions ${ \bf R} _i  \cap { \bf T} _d $ is the best response partition of ${ \bf T} _d $ induced by the game $(\zeta , { \bf M })$.

Suppose we have a generic best response game $(G,\zeta,\Phi ^0(G), { \bf M })$, where $G$ is a regular degree $d$ graph. This will be a regular automata
$(G,\zeta,\Phi ^0(G),F)$ where $F(D)$ is the strategy $j \in \zeta$ such that  \begin{equation} \frac{ { \displaystyle \sum_{i \in D} { \bf e} (i)  } } {d} \in {
\bf R} _j , \forall D \in \mathbb{P} ^d ( \zeta ). \end{equation}

\begin{figure}
      \centering
\includegraphics[scale=0.1]{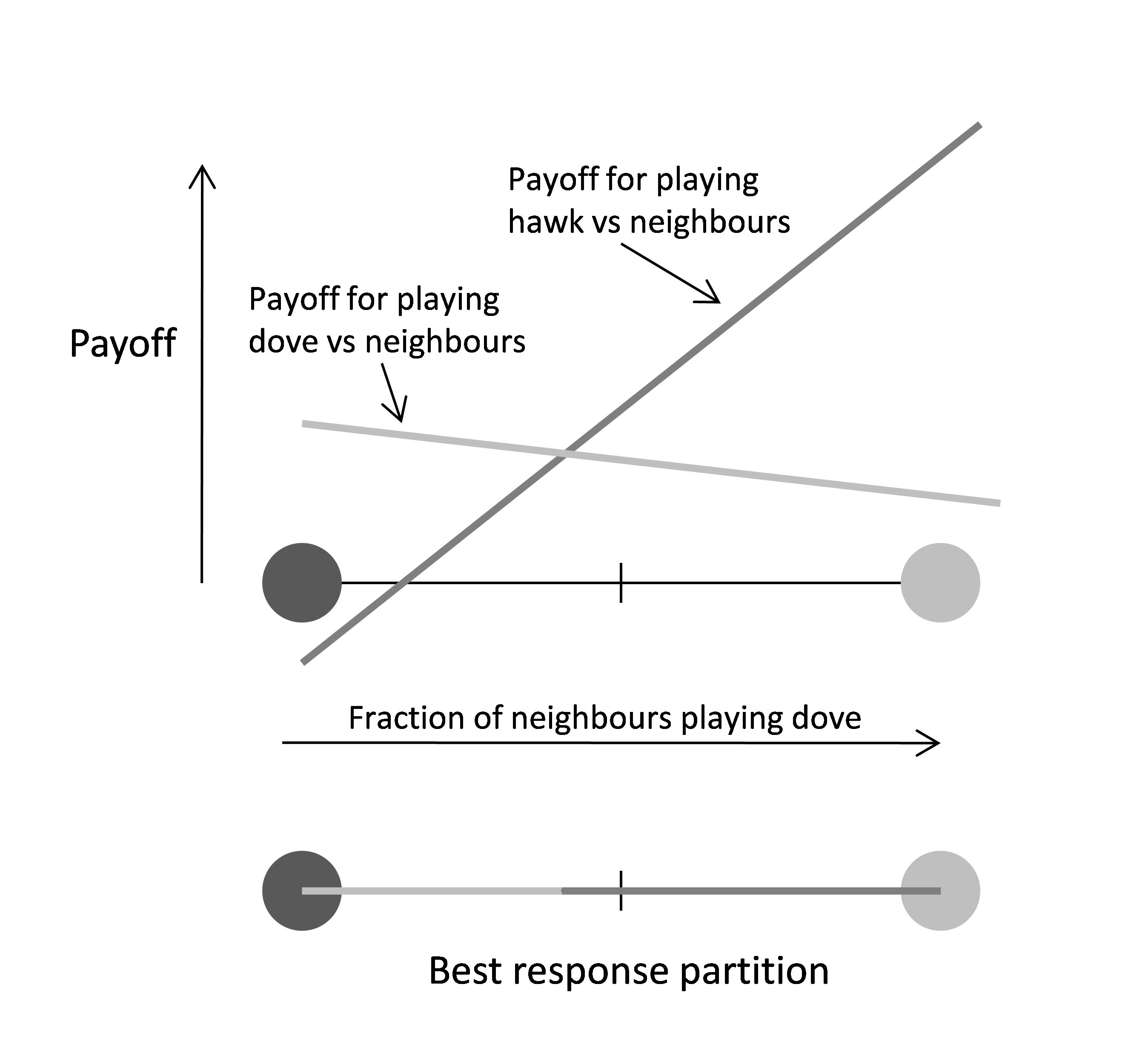}
      \caption{A plot showing the payoff received from using pure strategies against mixed strategies in the game described by equation \ref{hawkeqn}. The best
      response division is calculated by observing the pure strategy that scores best against each point of the strategy space}
      \label{hawkdoveplot}
   \end{figure}

Consider for example the Hawk-Dove game discussed in the previous section. This is a two strategy game so our strategy space $\Delta$ is the unit line. We can
plot the payoffs one receives from playing strategy $1$ or $2$ against the different strategy vectors ${ \bf x } \in \Delta$ as shown in figure
\ref{hawkdoveplot}. This induces a convex division of $\Delta$ into two best response regions ${ \bf R} _1 $ and ${ \bf R} _2 $. The update function $F$ is
determined by considering how the points of \begin{equation} { \bf T } _2  = \{ (1,0) , (1/2 , 1/2) , (0,1) \} \end{equation} are divided up into these best
response regions. For example $(1/2 , 1/2) \in { \bf R} _2 $ so $F(\{1 , 2\}) = 2$.

\begin{figure}
      \centering
\includegraphics[scale=0.1]{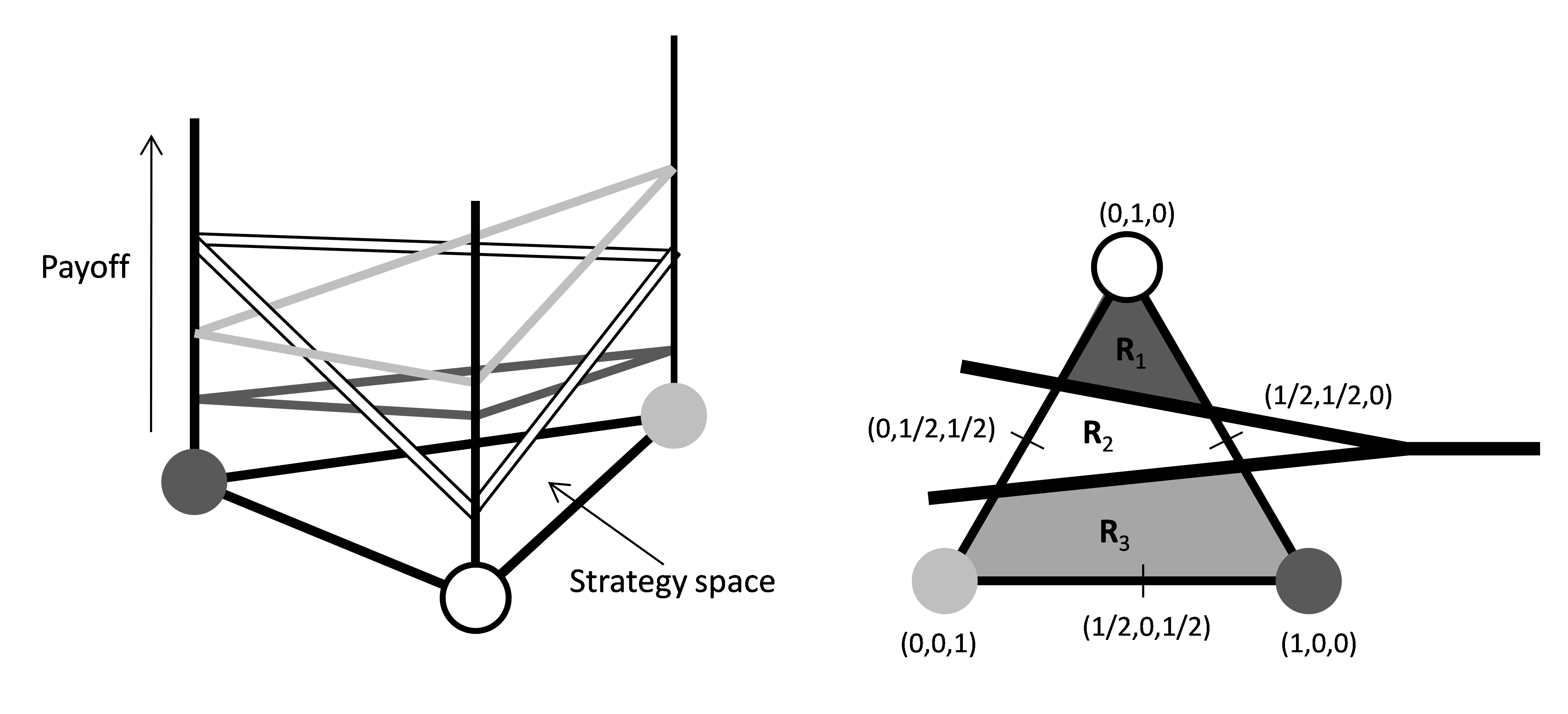}
      \caption{On the left is an illustration of how pure strategies score different payoffs against mixed strategies in three strategy games. On the right is
      the best response division associated with the game described by equation \ref{foresteqn}.}
      \label{threed}
   \end{figure}

For a 3 strategy game the simplex $\Delta$ is the unit triangle. Again we can plot the payoffs one receives from playing pure strategies against strategy vectors
in $ \Delta$. In figure \ref{threed} (left) the x-y plane represents the different strategy vectors, and the z coordinate representing the payoffs one receives
from employing different pure strategies against these vectors. Again this induces a division of the simplex $\Delta$ into convex best response regions.

\ \\

\begin{theorem} \label{bigbest}

$\forall d,k>0$

A partition of the points of ${ \bf T} _d $ into $m \leq k$ subsets ${ \bf W} _i  \subseteq { \bf T} _d $ is induced by the best response division associated
with some $k$ strategy game $( \zeta , { \bf M })$ if and only if the convex hulls of each pair of distinct subsets ${ \bf W} _i $ and ${ \bf W} _j $ do not
intersect.

\end{theorem}

\ \\

Theorem \ref{bigbest} allows us to determine when a given $k$ state regular automata is a best response game. The proof is in the appendix and the remainder of
this section describes the geometry of best response divisions in more detail. Given a non-singular payoff matrix ${\bf M}$
 we can find a division of the
$(k-1)$-dimensional space ${\bf S}$ into (open) best response regions ${\bf R} _i$. First we will consider best response divisions of ${ \bf S}$. Later we will
consider how such divisions divide up the points of the simplex of attainable strategies \begin{equation} \Delta = \{ {\bf x} \in { \bf S} : {\bf x}\ge 0 \}.
\end{equation} For each pair $i,j \in \{1,2,...,k\}: i \neq j$ let ${\bf H}(i,j) \subset {\bf S}$ be the set of all ${\bf x}\in {\bf S}$ where \begin{equation}
[{\bf M}{\bf x}]_i=[{\bf M}{\bf x}]_j. \end{equation} The hyperplane ${\bf H}(i,j)$ is the set where the payoffs to pure strategies $i$ and $j$ are equal. Since
${\bf M}$ is non-singular this hyperplane has dimension $k-2$. The hyperplane divides ${\bf S}$ into two regions, one where the payoff to pure strategy $i$
exceeds that to pure strategy $j$, and the other where the payoff to pure strategy $j$ exceeds that to pure strategy $i$. Each of the $\frac{n(n-1)}{2}$ pairs
$\{ i,j \}$ define such a dividing hyperplane. The set \begin{equation} \cup _{i , j \in \{ 1 , 2 ,..., k \} : i \neq j } {\bf H} (i,j) \end{equation} of all of
these hyperplanes together divide up the space ${ \bf S }$ into the $k!$ distinct regions corresponding to the distinct orderings of the $k$ payoffs at each
point. The best response region ${\bf R}_i$ is the union of $(k-1)!$ of these $k!$ regions, and these are necessarily contiguous. The set of regions correspond
to the orderings of $\{1,2,\ldots,k\}$, i.e. the permutations. The Cayley graph of the group ${\bf S}_k$ under the generator set of transposition of adjacent
elements corresponds to the adjacency of the ${\bf R}_i$. Each transposition corresponds to the crossing of a hyperplane where there is equality of the payoffs
for the elements which are transposed.

Alternately we can consider the sets \begin{equation} {\bf B}(i)=\{1,2,\ldots,k \} - \{ i \}, \end{equation} and ${\bf H(B}(i))$ is the set of ${\bf x} \in { \bf
S }$ where the payoffs to the elements of ${\bf B}(i)$ are equal. Since ${\bf M}$ is non-singular there exists a unique value \begin{equation} {\bf x} ^*
=\frac{{\bf M^{-1}}{\bf l}}{{\bf l^T}{\bf M^{-1}}{\bf l}} \end{equation} (A Nash equilibrium of the system). Each ${\bf H(B}(i))$ passes through ${ \bf x } ^*$
and on one side the payoff to $i$ is greater than that to all the others, and on the other side it is less. We consider the set of $k$ rays consisting of ${\bf
x}(k)$ and that part of ${\bf H(B}(i))$ where the payoff to $i$ is less than the payoff to the other strategies, ${\bf U}_i$. Since our matrix is non-singular
the set of rays form a basis for the simplex. For some set ${\bf V} \subset \{1,2,\ldots,k\}$ the convex combination of the corresponding rays, $\cup_{i \in
V}U_i$ is the set where there is equality of the payoffs to the set of payoffs indexed by the elements not in ${\bf V}$, all elements in ${\bf V}$ having lower
payoffs. The regions ${\bf R}_i$ are the interiors of the (closed) regions generated by convex combinations of points from ${\bf V}={\bf B}(i)$. Each of these
$k$ closed regions is bounded by $(k-1)$ rays.

Thus we have that the division of the unit simplex into the $k$ best response regions is simply achieved by taking a point in the hyperplane containing the unit
simplex, and $k$ rays emanating from that point with the condition that the reflection of any ray in the central point lies in the convex hull of the other
$(k-1)$ rays.

The converse of the above argument is simply that if we have a division of the hyperplane containing the unit simplex according to the above rule then we can
find a unique matrix which corresponds to those rays. Of course in the context of a game on a circle there will be many sets of rays which produce the same best
response regions, and thus many payoff matrices. Suppose then that we are given the specification of the rays as a set of linearly independent vectors ${\bf
u}_i$ for $i=1,2,\ldots,k$ and form the matrix ${\bf U}$ which has the columns equal to the ${\bf u}_i$, then we select any matrix ${\bf A}$ with $ith$ column
which have equal entries except for the diagonal entry which is smaller. Now we only require to find matrix ${\bf M}$ such that \begin{equation} {\bf MU}={\bf A}
\end{equation} i.e. ${\bf M}={\bf A}{\bf U}^{-1}$, for this matrix ${\bf M}$ to provide us with an appropriate payoff matrix, though we may require to add a
constant to all elements if we require payoffs to be positive.\\

Example. Suppose the rays from the central equilibrium value are such that the matrix ${\bf U}$ is given by \\

${\bf U}=
 \left(
  \begin{array}{cccc}
    1 & 2 & 3 & 3 \\
    2 & 3 & 4 & 1 \\
    3 & 1 & 2 & 4 \\
    4 & 4 & 1 & 2 \\
  \end{array}
\right) $

Note that the columns add to a constant but this is not required. Now we can select any appropriate matrix ${\bf A}$. For ease we take ${\bf -I}$, where ${\bf
I}$ is the unit matrix. We have

${\bf M}= \left(
  \begin{array}{cccc}
    0.5444 & -0.2333  & -0.3444 & -0.0111 \\
    -0.3889 & 0.1667 & 0.3889 & -0.2778 \\
    0.1111 & -0.3333 & -0.1111 & 0.2222 \\
    -0.3667 & 0.3000 & -0.0333 & -0.0333 \\
  \end{array}
\right) $

and we can add $1$ if we require positive entries.

\ \\

\section{Games on the circle with 2 or 3 strategies} \label{sec:twoandthree} \ \\

We can apply theorem \ref{bigbest} to enumerate the two strategy best response games on the circle. To do this we must simply list all the different possible
ways to divide up our simplex $\Delta$, the unit line, into two or less convex regions, with respect to the three points of ${ \bf T} _2 $ - the lines two end
points and the mid point.

There are only two ways of doing this, either all points belong to the same region or one end point belongs to one region and the other two points belong to the
other. We can take each of these two unlabeled divisions and apply labels to the regions, deciding which best response regions they represent. We hence find
that there are six non-identical two strategy best response game on the circle, three of which are permuationally distinct, meaning there are three fundamentally
different types of two strategy best response games (see Table \ref{3games}).

\ \\ \begin{table*}[http] \caption{The payoff inequalities describe the three types of two strategy game that induce fundamentally different dynamics in best
response games on the circle.} \begin{center} \begin{tabular}{l*{2}{c}r} Payoff inequalities that generate game type & Example game \\ \hline ${ \bf M }_ {1,1} >
{ \bf M }_ {2,1} $, ${ \bf M }_ {1,2} > { \bf M }_ {2,2}$  & Trivial  \\ ${ \bf M }_ {2,1} > { \bf M }_ {1,1} $, ${ \bf M }_ {2,1} + { \bf M }_ {2,2} > { \bf M
}_ {1,1} + { \bf M }_ {1,2} $, ${ \bf M }_ {1,2} > { \bf M }_ {2,2} $ & Hawk-Dove \\ ${ \bf M }_ {1,1} > { \bf M }_ {2,1} $, ${ \bf M }_ {2,1} + { \bf M }_ {2,2}
> { \bf M }_ {1,1} + { \bf M }_ {1,2} $, ${ \bf M }_ {2,2} > { \bf M }_ {1,2} $ & Stag Hunt \\ \end{tabular} \end{center} \label{3games} \end{table*}

The first type are games where one strategy strictly dominates. These systems induce very dull dynamics with every vertex constantly playing the dominating
strategy. Figure \ref{hawkdovespace} depicts the dynamics of a game of the second type. The dynamics induced correspond to Wolfram's automata number 95. When the
circle has even length there are two repelling fixed points, where no adjacent vertices share the same strategy. The system has many period two orbits which
quickly attract other configurations. The third type of game corresponds to Wolfram's automata number 160. When the circle has even length there is a repelling
period two orbit -jumping between the two configurations with no adjacent vertices sharing the same strategy. The system has many fixed points which quickly
attract other configurations.

We can use theorem \ref{bigbest} again to enumerate the best response games on the circle with three strategies. Recall how the best response division depicted
at the right of figure \ref{threed} induces the dynamics depicted in figure \ref{forestrand}. Our theorem implies that any division of $\Delta$ (which is the
unit triangle) into three or less convex regions is induced by some game. The update function induced by such a division depends upon the way the six points of
${ \bf T} _2$ (the 3 vertices and 3 edge-midpoints of the triangle) are partitioned into these best response regions.

To enumerate all of the three strategy games we must simply list all the fundamentally different ways of dividing up $\Delta$ into three or less convex regions
with respect to the points of ${ \bf T} _2$. There are 12 fundamentally different ways to perform such a division.

Each division $p$ induces an equivalence class, which is the set of best response divisions of ${ \bf T} _2$ which can be attained by taking $p$ and labeling
the regions with different strategies -deciding which best response region each region of $p$ represents. By looking at the different labellings of the 12
divisions we find that there are $285$ non-identical three strategy best response games on the circle, 52 of which are permutationaly distinct. We give space
time plots of the 52 cases in the appendix (subsection \ref{subsec:3st}), together with diagrams that show the divisions corresponding to the equivalence
classes of the games.

\ \\

\section{Games with more strategies} \label{bestgen} \ \\

Enumeration of best response games on the circle with more than three strategies is difficult to do in the same visual manner as above. The reason is that the
simplex is high dimensional and the number of differenr convex divisions of the simplex with respect to ${ \bf T } _2$ is large. Since the set of $k$ strategy best response games on the circle are a subset of the set of $k$ state regular automata one fruitful question to ask is \emph{when is a $k$ state regular automata on
the circle not a $k$ strategy best response game ?}

We can think of each regular automata, $(G,S, \Phi ^0(G),F)$, on a $d$ regular graph $G$, as inducing a partition of ${ \bf T} _d  \subset \Delta$ in a similar
way to the way we did for best response games. To do this is that we think of our set of states as numbers $S=\{ 1,2,...,k \}$ and we think of each $D \in
\mathbb{P} ^d (S)$ as a point \begin{equation} {\bf P} (D) = \frac{{ \displaystyle \sum_{i \in D} { \bf e} (i) } } {d} \in { \bf T} _d \end{equation} in the
simplex. We think of the points ${ \bf T} _d = \{ {\bf P} (D) : D \in \mathbb{P} ^d (S) \}$ as being partitioned into $m \leq k$ subsets ${ \bf W} _i $ where ${
\bf W} _i $ is the set of all points ${\bf P} (D)$ such that $F(D)=i$.

The converse of theorem \ref{bigbest} is that a regular automata $(G,S, \Phi ^0(G),F)$ is not a $k$ strategy best response game if and only if the partition of
${ \bf T_d }$ that $(G,S, \Phi ^0(G),F)$ induces has a distinct pair of sets ${ \bf W} _i $ and ${ \bf W} _j $ with intersecting convex hulls.

So to answer our question, we should find all of the pairs of disjoint subsets $X, Y \subseteq \mathbb{P} ^d (S)$ such the convex hulls of $\{ {\bf P} (D) | D
\in X \}$ and $\{ {\bf P} (D) | D \in Y \}$ intersect. We call such an $X,Y$ pair $(k,d)$ unacceptable because a $k$ state regular automata $(G,S, \Phi ^0(G),F)$
on a $d$ regular graph is not a best response game if and only if $S$ has a pair of states $i \neq j$, such that $F^{-1} (i) , F^{-1} (j)$ are $(k,d)$
unacceptable.

Clearly if a pair $X, Y \subseteq \mathbb{P} ^d (S)$ are such that there is a pair $X' \subseteq X$ and $Y' \subseteq Y$ where $X',Y'$ are $(m,d)$ unacceptable,
for $m \leq k$, then $X,Y$ are $(k,d)$ unacceptable. Knowing this we can tighten the definition of unacceptable pairs, to lessen the number of objects we need to
catalogue to determine whether or not a regular automata is a best response game.

We say that a pair $X, Y \subseteq \mathbb{P} ^d (S)$ are fundamentally $(k,d)$ unacceptable if and only if $X,Y$ are $(k,d)$ unacceptable and $\forall X'
\subseteq X$, $\forall Y' \subseteq Y$, $\forall m \leq k$ we have that $X',Y'$ are $(m,d)$ unacceptable implies $\{ X' , Y' \} = \{ X , Y \}$ and $m=k$.

In other words a fundamentally unacceptable pair is an unacceptable pair that properly contains no other unacceptable pairs.  So we arrive at theorem
\ref{unacceptable}.

\begin{theorem} \label{unacceptable}

$\forall d,k>0$

A $k$ state regular automata $(G,S, \Phi ^0(G),F)$ on a $d$ regular graph is a best response game if and only if $\forall m \leq k$, for every pair of states $i
\neq j$ of $S$, there does not exist a pair $X \subseteq F^{-1} (i)$,$Y \subseteq F^{-1} (j)$ that is fundamentally $(m , d)$ unacceptable.

\end{theorem}

Our enumeration problem is hence transformed into the problem of finding the set of permuationally distinct fundamentally unacceptable pairs. The set of
different convex partitions of ${ \bf T } _d$ can be found by listing all the permuationally distinct partitions of ${ \bf T } _d$ and then filtering out those
partitions which involve a pairs $X,Y$ such that $X' \subseteq X$, $Y' \subseteq Y$ is fundamentally $(m,d)$ unacceptable, for $m \leq k$.

Let us consider the problem on the circle, when $d=2$. There are no fundamentally $(1,2)$ unacceptable pair because $\mathbb{P} ^2 ( \{1 \} )$ cannot be split
into two disjoint non-empty sets. The fundamentally $(2,2)$ unacceptable pairs can be found visually, the only permuationally distinct way one may choose two
disjoint subsets ${ \bf A }$ and ${ \bf B }$ of ${ \bf T_2 } = \{ (1,0) , (1/2,1/2) , (0,1) \}$ such that the convex hulls of ${ \bf A }$ and ${ \bf B }$
intersect is ${ \bf A } = \{ (1,0) , (0,1) \}$ and ${ \bf B } = \{ (1/2,1/2) \}$. The pair $X,Y$, where $X = \{ \{1,1 \} , \{2,2 \} \}$ and $Y = \{ \{1,2 \} \}$
is hence the only permuationally distinct fundamentally $(2,2)$ unacceptable pair.

\begin{figure}
      \centering
\includegraphics[scale=0.1]{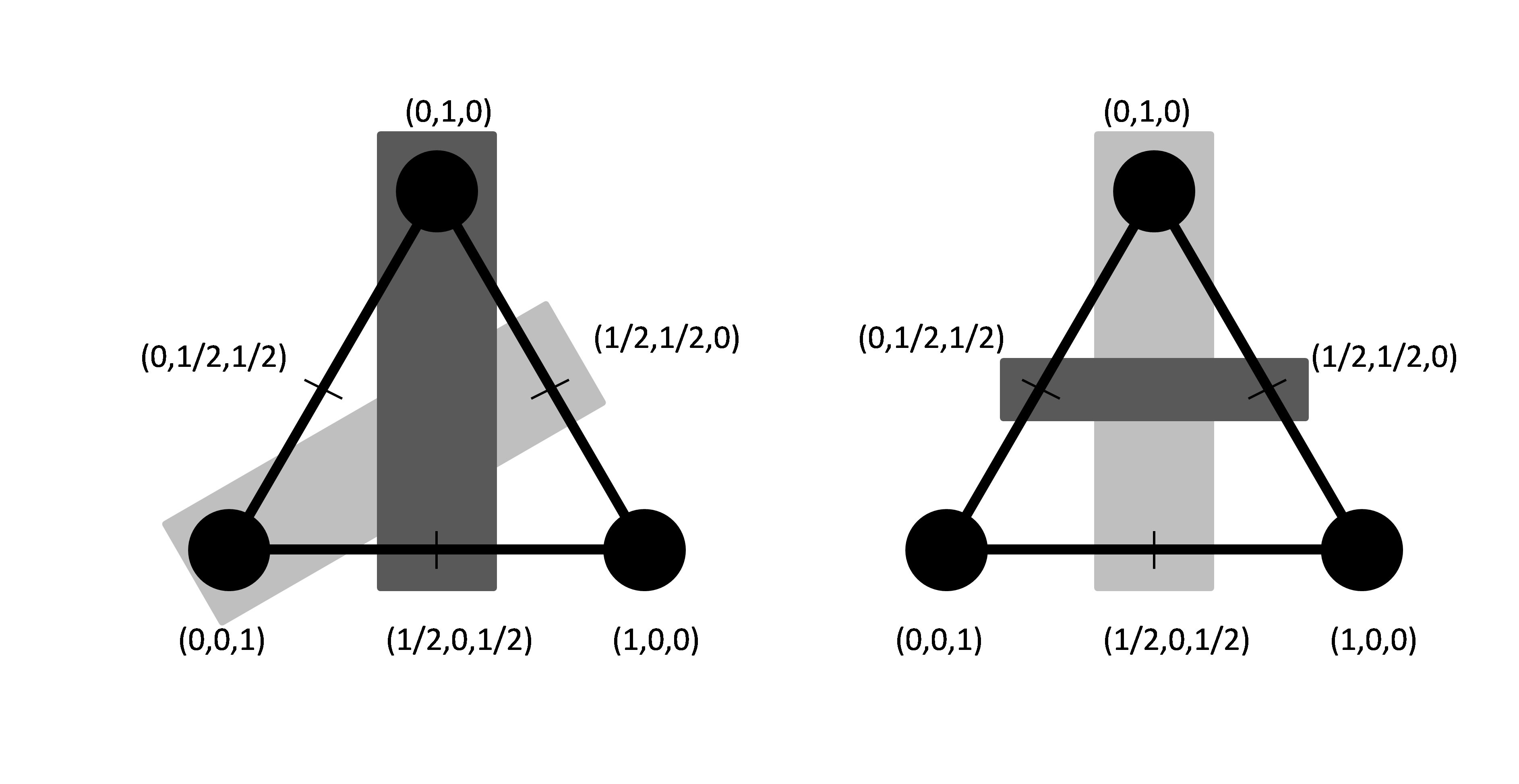}
      \caption{The two fundamentally $(3,2)$ unacceptable pairs within the two dimensional simplex $\Delta$. The left shows $\{ \{2,2\} , \{1 , 3 \} \}$, $\{ \{
      3 , 3 \}, \{ 1 , 2 \} \}$, the right shows $ \{ \{ 1 , 2 \}, \{ 2 , 3 \} \}$, $\{ \{ 2 , 2 \}, \{1 , 3 \} \}$.}
      \label{u3}
   \end{figure}

The set of fundamentally $(3,2)$ unacceptable pairs can again be found visually. It is easy to see that, if the convex hulls of two disjoint sets of ${ \bf T}
_2$, in the unit triangle, intersect, then one of the two situations depicted in figure \ref{u3} must have occurred.

For a pair of disjoint sets $X,Y \subset \mathbb{P} ^ 2 (S)$, let $Gr(X,Y)$ be the graph with a vertex set consisting of all $x \in S$ such that $x$ is a member
of a pair in $X$ or $Y$, and edge set consisting of dark gray edges $X$ and light gray edges $Y$. An alternating walk on such a graph $Gr(X,Y)$ is a walk on the
edges of $Gr(X,Y)$ such that every edge traversed is a different colour to the previously traversed edge. An alternating cycle of such a graph is an alternating
walk that finishes on the same vertex where it started -returning along an edge of a different colour to the colour of the edge that the walk first traversed.

\begin{figure}
      \centering
\includegraphics[scale=0.1]{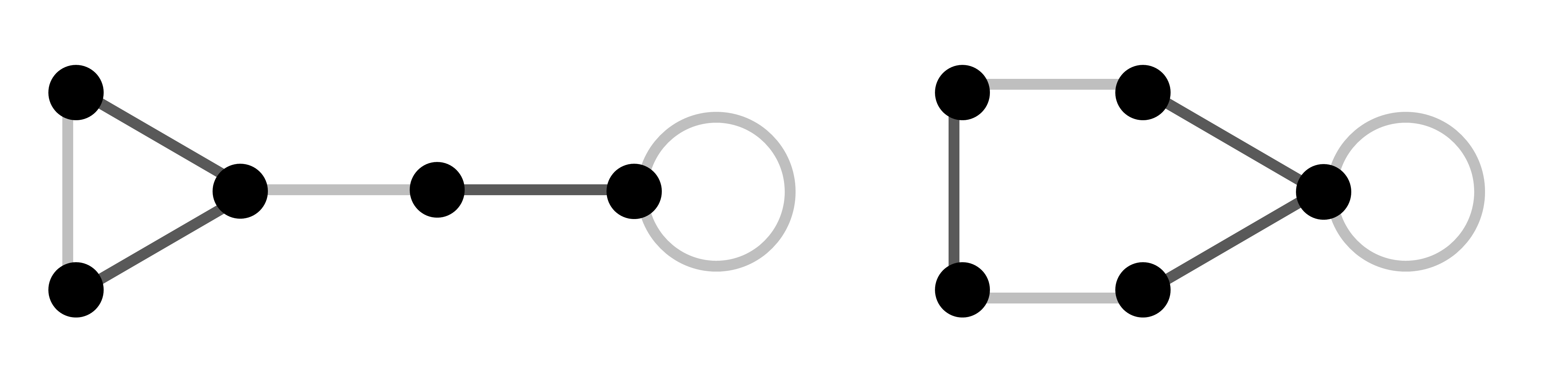}

      \caption{On the left is an illustration of $Dum(2,0)_5$. On the right is an illustration of $Dum(5,0)_5$. Both graphs have been given a good colouring.
}
      \label{dumbegg}
   \end{figure}

\begin{lemma} \label{cyc}

A pair $X,Y$ is $(k,2)$ unacceptable if and only if $Gr(X,Y)$ has an alternating cycle.

\end{lemma}

\ \\ This leads to a result that allows us to completely characterise the set of fundamentally $(k,2)$ unacceptable graphs for generic $k$. Recall that $C_n$
denotes the $n$ vertex circle graph, let $C_1$ be a single vertex with a self loop. Let a $k$ vertex dumbbell graph $Dum(a,b)_k$, where $a+b < k$, be the fusing
of two circle graphs $C_{a+1}$ and $C_{b+1}$ to the two end points of a line graph (by identifying/ overlapping vertices) so that the resulting graph,
$Dum(a,b)_k$, has $k$ vertices (see figure \ref{dumbegg}). Note that when $a+b = k-1$, the connecting line between the two circles in $Dum(a,b)_k$ has no edges,
and hence $Dum(a,b)_k$ resembles a figure 8 in that it consists of two circles intersecting at one vertex.

Let a good colouring of a graph $G$ be a colouring of its edges with dark gray and light gray such that, if a vertex $v \in G$ only has two edges incident on it
then the two edges are painted different colours and otherwise two edges incident on a vertex $v$ are painted different colours if and only if they do not lie on
the same cycle of $G$.

\begin{theorem} \label{fundun}

 $X,Y$ is fundamentally $(k,2)$ unacceptable if and only if one of the following conditions hold;

1) $k=2$ and $Gr(X,Y)$ is a good colouring of $Dum(0,0)_2$.

2) $k>2$ is even and $Gr(X,Y)$ is a good colouring of $C_k$ or $Dum(a,b)_k$ where $a,b \in \{ 0, 1 ,...,k \}$ are even and such that $a+b <k$

3) $k>2$ is odd and $Gr(X,Y)$ is a good colouring of $Dum(a,b)_k$ where $a,b \in \{ 0, 1 ,...,k \}$ are even and such that $a+b <k$.

\end{theorem}

Using theorems \ref{unacceptable} and \ref{fundun} we can make an algorithm to check if a regular automata on the circle corresponds to a best response game and
hence we can solve the problem of finding all of the fundamentally different $4$ strategy best response games on the circle. The way we do this is to use a
computer to generate the set of all four state degree 2 regular automata and then filter this set, removing those rules that do not correspond to best response
games. We find that there are 143524 non-identical four strategy games and 6041 permutationally distinct games.

\ \\ \section{Games on other graphs} \label{css}

\ \\

When dealing with degree three graphs, the different update functions $F$ that can occur correspond to convex divisions of $\Delta$ with respect to the points
of ${ \bf T } _3$. With two strategies, we may enumerate the possible best response games by listing the different divisions of the unit line $\Delta$ into
$\leq 2$ convex regions with respect to the points of ${ \bf T} _3 = \{ (1,0) , (1/3, 2/3), (2/3,1/3) , (0,1)\}$. Using this approach one can determine the 5
permutationally distinct update functions $F$ that could be induced by two strategy best response games on degree three graphs. Its important to note that the
update rules found in this way could be evolved upon many different graph topologies. One could consider dynamics of the cube, the Peterson graph or any other
degree three graph. The circle with self-linkage is the degree three graph obtained by taking a circle and linking each vertex to itself. Looking at best
response games on the circle with self linkage is beneficial because the resulting one dimensional cellular automata can be visualised using space time plots.
The permutationally distinct two strategy best response games running on the circle with self linkage correspond to rules Wolfram's elementary cellular automata
numbers 0, 23, 127, 128 and 232. One may enumerate the different three strategy games on degree three graphs in a similar manner by listing the different ways to
cut up the unit triangle into convex regions with respect to the points of ${ \bf T} _3$. Using this method one finds that there are 82 fundamentally different
three strategy best response games on degree three graphs.

These methods can be applied to enumerate the number of $k$ strategy games on degree $d$ graphs. Such an enumeration seems difficult to do for generic $k$ and
$d$. Theorem \ref{bigbest} provides a way to do such an enumeration in theory but with no result like theorem \ref{fundun} (which allows us to quickly filter out
unviable regular automata) the computation would be slow for $d>2$.

Our results can be extended to deal with non-regular graphs. Suppose we have a graph $G$ and $\{ d_i : 1 \leq i \leq n \}$ is the set of all $d_i$ such that
there is a vertex of $G$ with degree $d_i$. To enumerate the different best response games on $G$ one must simply list all the different ways to devide
$\Delta$ into $k$ or less convex regions with respect to the points of $\cup _{i=1} ^n { \bf T} _{d_i}$.

\ \\

\section{Appendix}\ \\

\subsection{Proof of theorem \ref{bigbest}}

Any game $(G,{ \bf M })$ will induce a division of the extended strategy space ${ \bf S }$ into best response regions ${ \bf R } _i$. To show these best
response region are convex, consider two points ${ \bf x }$ and ${ \bf y }$ within ${ \bf R} _i$, then $[{ \bf Mx }]_i > [{ \bf M x}]_j$ and $[{ \bf My }]_i > [{
\bf My }]_j$, $\forall j \in \{1,2,...,k\} - i$, by definition. Since ${ \bf M }$ is a linear mapping any convex combination $\lambda { \bf x } + (1- \lambda ) {
\bf y }$, for $\lambda \in [0,1]$ will be such that $[{ \bf M } ( \lambda { \bf x } + (1- \lambda ) { \bf y } ) ]_i > [{ \bf M } ( \lambda { \bf x } + (1-
\lambda ) { \bf y } ) ]_j$ , $\forall j \in \{1,2,...,k\} - i$. This means $\lambda { \bf x } + (1- \lambda ) { \bf y }$ also lies within the best response
region ${ \bf R }_i$. So every best response region ${ \bf R}_i$ is convex. This means our game induces partition of ${ \bf T} _d$ into best response regions
${ \bf W } _i = { \bf R} _i \cap { \bf T} _d$, such that the convex hulls of any two sets ${ \bf W} _i \neq { \bf W} _j$ do not overlap.

Proving the converse is more involved.

Suppose we have a partition of the points of ${ \bf T } _d$ into $m \leq k$ sets ${ \bf W } _i$ such that the convex hulls of each pair of sets do not overlap.
There will be a family of appropriate divisions of ${ \bf S } \simeq \mathbb {R} ^{k-1}$, into $k$ convex open sets ${ \bf P } _i$, that generate such a
partition of ${ \bf T } _d$ in that $\forall i$, ${ \bf W } _i = { \bf P} _i \cap { \bf T } _d$.

Each such division, where every region ${ \bf P } _i$ has non zero volume, must be generated by a set of dividing hyperplanes, which is a set of $k-2$
dimensional hyperplanes that cut up the space into different regions. Each ${ \bf P } _i$ is a polyhedral set and every $k-2$ dimensional face of ${ \bf P } _i$
is the intersection of the closure of ${ \bf P } _i$ with one of its neighboring regions. The set of dividing hyperplanes which generates such a division is the
set of affine hulls of all such faces of all regions.

Among our family of appropriate divisions there will be a division of ${ \bf S}$ into $k$ non-zero volume, convex sets ${ \bf P}_i$ with the property that each
set of $k-1$ dividing hyperplanes involved in this division will meet at a single point, we will call such a division \emph{proper}. It is a well known result
that almost every arrangement of $k-1$ hyperplanes of dimension $k-2$ in $\mathbb{R} ^{k-1}$ will have a common point, such a point will always exist provided no
two of these hyperplanes have parallel subspaces. Any division can be made proper by doing an infinitesimal perturbation of the positioning of the dividing
hyperplanes involved. Since the points of ${ \bf T } _d$ are distantly spaced such a perturbation will not effect the way ${ \bf T } _d$ is partitioned up. This
means an appropriate proper division exists.

Suppose ${ \bf P} _i$ is a region within an appropriate proper division. We shall use a proof by contradiction to show that ${ \bf P} _i$ has a finite extreme
point (a vertex).
 Suppose (falsely) that ${ \bf P} _i$ does not have a finite extreme point. Let $\overline{{ \bf X}}$ denote the closure of ${ \bf X}$. Any closed convex set,
 like $\overline{{ \bf P} _i}$, with no finite extreme point, must contain a line ${ \bf L }$ (extending infinitely in both directions). Any translation of ${
 \bf L}$ that intersects with $\overline{{ \bf P} _i}$ must also be contained within $\overline{{ \bf P} _i}$.
Let ${ \bf P} _j$ be a region adjacent to ${ \bf P} _i$. Any translation of ${ \bf L }$ that intersects with $\overline{{ \bf P} _i} \cap \overline{{ \bf P} _j}
$ must be contained within $\overline{{ \bf P} _i} \cap \overline{{ \bf P} _j} $. This means any translation of ${ \bf L}$ that intersects $\overline{{ \bf P}
_j}$ must be contained within $\overline{{ \bf P} _j}$. This argument can be continued to show that every region contains a translation of ${ \bf L}$ and every
dividing hyperplane contains a translation of ${ \bf L}$.  This contradicts our assumption that the division is proper because such an arrangement of dividing
hyperplanes cannot meet at a point. Every $k-2$ dimensional cross section of our hyperplane arrangement attained by slicing perpendicular to ${ \bf L}$ will look
the same (irrespective of how far along ${ \bf L}$ one chooses to slice) so there cannot be a point where all the dividing hyperplanes meet. This contradiction
implies every region ${ \bf P }_i$ must have a vertex.

Since ${ \bf P }_i$ is $k-1$ dimensional a vertex of ${ \bf P }_i$ must be the intersection of at least $k-1$ of its faces. Each of ${ \bf P }_i$'s faces is
$\overline{{ \bf P} _i} \cap \overline{{ \bf P} _j} $ for some neighbouring region ${ \bf P} _j$. There are only $k$ regions so ${ \bf P }_i$ can have at most
$k-1$ faces. Hence ${ \bf P }_i$ has just one vertex ${ \bf v }$, and ${ \bf v }$ is the intersection of the closures of all $k$ regions. Let ${ \bf I} (i)$ be
the intersection of the closures of every region except ${ \bf P } _i$, it follows that ${ \bf I } (i)$ will be a one dimensional ray that is a common one
dimensional edge of every region except ${ \bf P }_i$. There will be $k$ such one dimensional rays ${ \bf I } (i)$, that all meet at ${ \bf v }$ and every region
${ \bf P } _j$ will be the interior of the convex hull of $\{ { \bf I } (a) : a \in \{1,2,...,k\} - j \}$. Each ray must lie outside of the convex hull of the
other $k-1$ rays (otherwise the interior of two regions would intersect and we would not have a division). An equivalent way to say this is that the reflection
of any ray in ${ \bf v} $ lies within the convex hull of the other $k-1$ rays.

Since our regions meet at a central point with $k$ emanating rays (that meet the appropriate conditions) we can use the results from section \ref{sec:convgeo} to
construct a non-singular payoff matrix ${ \bf M }$ which generates our convex division. Under the game with payoff matrix ${ \bf M }$ the $i$th best response
region ${ \bf R} _i$ will be equal to the convex region ${ \bf P} _i$, $\forall i \in \{1,2,...,k\}$. $\Box$

%Thus the division of the unit simplex into the $k$ best response regions can be generated by taking a point in the hyperplane containing the unit
%simplex, and $k$ rays emanating from that point which are such that the reflection of any ray in the central point lies in the convex hull of the other
%$(k-1)$ rays. The division of the unit simplex into the $k$ best response regions is obtained by identifying the $i$th best response region with the convex hull of all rays except the $i$th ray.

%The converse of the above argument is simply that if we have a division of the hyperplane containing the unit simplex according to the above rule then we can
%find a unique matrix which corresponds to those rays.

\subsection{Proof of lemma \ref{cyc}}

We will show that a pairs unacceptability implies the presence of an alternating cycle. Suppose that $X,Y \subseteq \mathbb{P} ^2 (S) $ is a $(k,2)$ unacceptable
pair, then by definition, there must exist subsets $X' \subseteq X$, $Y' \subseteq Y$ and sets of positive reals $\{ \lambda _ { \{a,b \} } >0 : \{a,b \} \in X'
\}$ , $\{ \mu _ { \{a,b \} } >0 : \{a,b \} \in Y' \}$ such that \begin{equation} { \displaystyle \sum _ { \{a,b \} \in X' } \lambda _{ \{ a,b \} } } = {
\displaystyle \sum _ { \{a,b \} \in Y' } \mu _{ \{ a,b \} } } =1 \end{equation} \ \\
 and
\begin{equation}{ \displaystyle \sum _ { \{a,b \} \in X' } \lambda _{ \{ a,b \} } ({ \bf e } (a) + { \bf e } (b))/2 } = { \displaystyle \sum _ { \{a,b \} \in Y'
} \mu _{ \{ a,b \} } ( { \bf e } (a) + { \bf e }(b))/2 }. \end{equation} \ \\ Now consider the graph $Gr(X' , Y')$ with each dark gray edge $\{a, b \} \in X'$
weighted with the constant $ \lambda _{ \{ a,b \} }$ and each light gray edge $\{a, b \} \in Y'$ weighted with the constant $ \mu _{ \{ a,b \} }$. The sum of the
weights of the dark gray edges incident upon any vertex will be equal to the sum of the weights of the light gray edges that are incident upon that vertex (where
self edges are counted as being incident twice). Suppose $w$ is the minimal weight on any edge of $Gr(X' , Y')$, let us multiply all of the weights of $Gr(X' ,
Y')$'s edges by $3/w$, so that all of the weights will be at least $3$.

Now start on any vertex of $Gr(X',Y')$, and walk along a dark gray edge, when the walk traverses an edge, reduce the weight of that edge by 1. After traversing a
dark gray edge, let the walk traverse a light gray edge, then a dark gray edge, then a light gray... and continue in this manner, reducing the weight of every
traversed edge. When an edge reaches weight $\leq 0$ it disappears and can no longer be used.

Every vertex must have at least two incident edges -one of each colour and such a walk is allowed to traverse each edge at least twice. Moreover, every time the
walk approaches a vertex with an edge of one colour, it will be able to leave the vertex with an edge of the other colour (at least this will be true until an
edge has been traversed twice). Clearly such a walk will be allowed to continue, in an alternating manner, until an edge is traversed three times. After an edge
has been traversed three times it follows that some vertex $v$ must have been visited three times. This implies that an alternating cycle has been generated. To
see this suppose, without loss of generality, that our walk first leaves $v$ along a dark gray edge. If the walk returns to $v$, for the first time, along a
light gray edge then an alternating cycle has clearly been generated. If, on the other hand, the walk returns to $v$, for the first time along a dark gray edge
then it must leave $v$, for the second time, along a light gray edge. When the walk returns to $v$ for the second time it will complete an alternating cycle. To
see this note that whatever the colour of the edge which the walk uses to return to $v$ for the second time, the walk will have used an edge of the opposite
colour to leave $v$ previously. This shows a pairs unacceptability implies the presence of an alternating cycle.

To see the converse suppose that the graph $Gr(X,Y)$ contains an alternating cycle $Gr(X',Y')$ with $X' \subseteq X$, $Y' \subseteq Y$. Now $\forall \{a,b \} \in
X'$ let $\lambda _ { \{ a , b \} }$ be the number of times that the edge $\{a,b \}$ is traversed in the alternating cycle $Gr(X',Y')$. Similarly $\forall \{a,b
\} \in Y'$ let $\mu _ { \{ a , b \} }$ be the number of times that the edge $\{a,b \}$ is traversed in the alternating cycle $Gr(X',Y')$. We refer to $\lambda _
{ \{ a , b \} }$ as the weight of the dark gray edge $\{a,b \} \in X'$ and we refer to $\mu _ { \{ a , b \} }$ as the weight of the light gray edge $\{a,b \} \in
Y'$.

Our alternating cycle will be such that the number of traversals of dark gray edges must be equal to the number of traversals of light gray edges, and hence our
coefficients will be such that \begin{equation} { \displaystyle \sum _ { \{ a , b \} \in X' } \lambda _ { \{ a , b \} } } = { \displaystyle \sum _ { \{ a , b \}
\in Y' } \mu _ { \{ a , b \} } } = I, \end{equation} \ \\ for some constant $I>0$.

\ \\
 The alternating cycle will be a walk such that every time a vertex is approached along an edge of one colour the walk will leave the vertex along an edge of
 another colour and each edge $\{a,b \}$ of $Gr(X',Y')$ is traversed by this walk a number of times equal to its weight. It follows that, for every vertex $v$ of

 $Gr(X',Y')$, the sum of the weights of $v$'s incident dark gray edges is equal to the sum of the weights of $v$'s incident light gray edges (where self edges
 are counted as being incident twice).

 Hence we get \begin{equation} { \displaystyle \sum _ { \{a,b \} \in X' } \lambda _{ \{ a,b \} } ({ \bf e }(a) + { \bf e }(b))/2 } = { \displaystyle \sum _ {
 \{a,b \} \in Y' } \mu _{ \{ a,b \} } ( { \bf e } (a) + { \bf e } (b))/2 }, \end{equation}
\ \\ so we can divide all of our parameters $\lambda _{ \{ a,b \}}$ and $\mu _{ \{ a,b \} }$ by our constant $I$ to get the set of convex coefficients which
describe a point where the convex hull of $\{ {\bf P} (D) | D \in X \}$ and $\{ {\bf P} (D) | D \in Y \}$ intersect. $\Box$

\subsection{Proof of theorem \ref{fundun}}

Suppose $X,Y$ is fundamentally $(k,2)$ unacceptable, then according to the definition of fundamentally unacceptable pairs and lemma \ref{cyc}, $Gr(X,Y)$ is an
alternating cycle, and hence must be connected. Moreover there can only be one recolouring of the edges of $Gr(X,Y)$, then is an alternating cycle (that
recolouring which just swaps the colours of every edge). If this were not so then $Gr(X,Y)$ would contain more than one fundamentally different alternating
cycle, and hence would not be fundamentally unacceptable.

Now suppose that $Gr(X,Y)$ has an even cycle $C$ on more than three vertices. $C$ can be recoloured to be an alternating cycle, and this means that $Gr(X,Y)$
consists of exactly $C$ and nothing more. Note that $C$ is a good colouring a circle graph on an even number of vertices.

Next suppose that $Gr(X,Y)$ has no even cycles, and at most one odd cycle. In this case $X,Y$ can not be fundamentally $(k,2)$ unacceptable. To see this consider
a walk which is an alternating cycle. Such a walk must traverse a cycle of the graph. The walk cannot take place on a purely linear graph (i.e. a line graph)
because this would imply that the walk must change direction at some point -back tracking along the edge just used, but this violates our requirement that the
colours of edges used alternate. Now let us (falsely) suppose that our $Gr(X,Y)$ does have a walk which is an alternating cycle. Since our walk is required to
traverse a cycle of $Gr(X,Y)$ we can assume (without loss of generality) that the walk begins at a vertex $v$ on the odd cycle of $Gr(X,Y)$ and immediately
traverses the cycle. When the walk returns to $v$, for the first time, it will do so along an edge of the same colour as the first edge traversed in the walk. To
complete an alternating cycle the walk must return to $v$ along a different colour. Clearly traversing the odd cycle again is not going to achieve this. The only
other way to try (in our efforts to form an alternating cycle) is to have the walk leave the odd cycle, to visit other vertices of $Gr(X,Y)$. This cannot be done
however because $Gr(X,Y)$ only holds one cycle. Once our walk leaves this cycle it will have no way to return except to backtrack, which we have already shown is
not allowed.

Now the only other possible case is that $Gr(X,Y)$ contains no even cycles and at least two odd cycles $C'$ and $C$. Since $Gr(X,Y)$ is connected there must be a
linear path $P$ (a sequence of end to end edges forming a line graph) between $C'$ and $C$. Now $C' \cup P \cup C$ can be given an edge recolouring (the good
colouring of $C' \cup P \cup C$) that \emph{is} an alternating cycle. We shall construct such a cycle by describing a walk (it will be clear that $C' \cup P \cup
C$'s edges can be coloured in such a way that the edge colours alternate on this walk).  Suppose our walk starts off at the intersection of $P$ and $C'$ (the
vertex $e'$ of $C'$ which is an point of the line graph $P$). Suppose the walk begins by traversing $C'$, starting off with a dark gray edge. After traversing
$C'$, the walk will return to $e'$ along a dark gray edge. Next the walk travels along a light gray edge of $P$ towards $C$. Suppose the walk continues traveling
along $P$ until it reaches the other end point, $e$, of $P$ (which intersects with $C$). The walk then moves around $C$, returning to $e$ on the same colour edge
by which it set off (on $C$), and then the walk travels back along $P$, to $e'$. When the walk returns to $e'$ it will do so along a light gray edge, thus
completing the alternating cycle. So we have shown that if $Gr(X,Y)$ contains more than one odd cycle, then $Gr(X,Y)$ must exactly be of the form $C' \cup P \cup
C$, which is exactly the form of a dumbbell graph $DUM(a,b)_k$ where $a,b \in \{ 0, 1 ,...,k \}$ are even and such that $a+b <k$. Such a graph will only have one
fundamentally different alternating cycle, which can be found by doing a good colouring of it.

So we have shown that all the graphs associated with fundamentally $(k,2)$ unacceptable pairs $X,Y$ lie in the set $\Omega _k$ of graphs described in the theorem
(even length circle graphs and dumbbell graphs with odd cycles). All that remains is to show that there are not any graphs within this set that are not $(k,2)$
unacceptable. We know that each of these graphs has at most one fundamentally different alternating cycle and no unnecessary extra structure, so all that is left
is to show that no graph in $\Omega _k$ has a proper subgraph that is a member of $\Omega _m$ for $ m \leq k$. For a dumbbell graph with two odd cycles, this is
obvious since every proper subgraph of it is neither a dumbbell graph, nor a circle graph of even length. Similarly no proper subgraph of a circle graph is a
circle graph or a dumbbell graph. $\Box$

\ \\

\subsection{The different three strategy games on the circle} \label{subsec:3st} In this subsection we give example space time plots (from random initial
conditions) showing the dynamics of each of the 52 non-identical best response games on the circle (see section \ref{sec:twoandthree}). We group these plots
together with the diagrams that show the unlabeled partitions of ${\bf T} _2$ which can be coloured to yield their best response partitions.

\ \\

\includegraphics[scale=0.7]{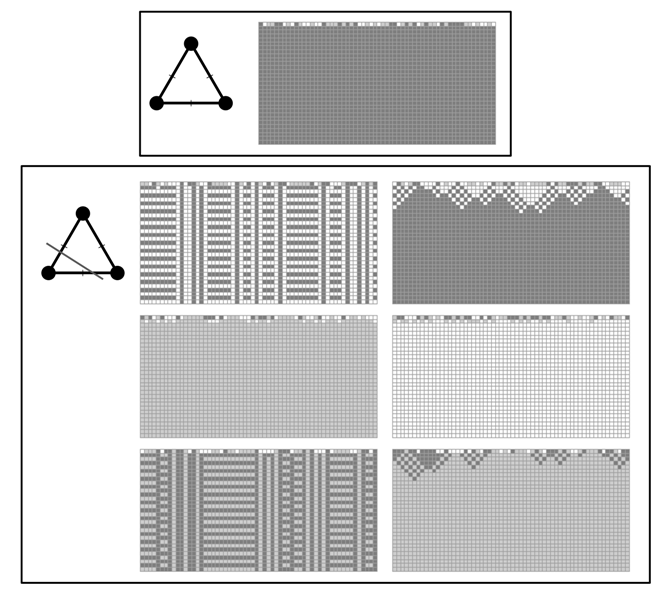}

\includegraphics[scale=0.7]{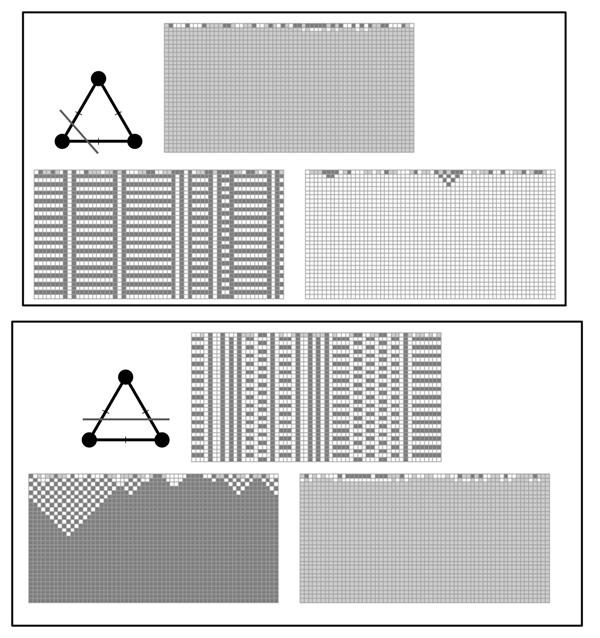}

\includegraphics[scale=0.7]{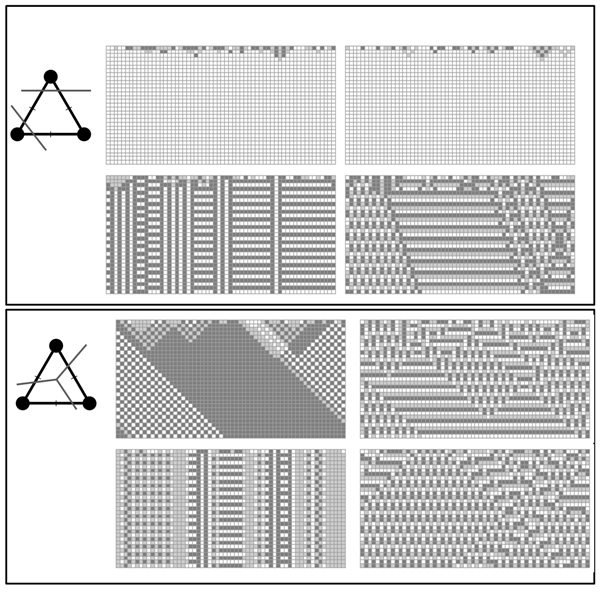}

\includegraphics[scale=0.7]{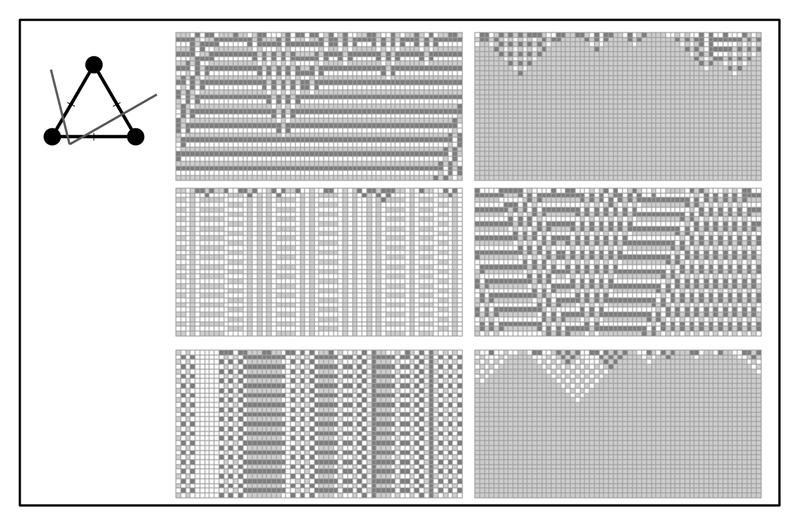}

\includegraphics[scale=0.7]{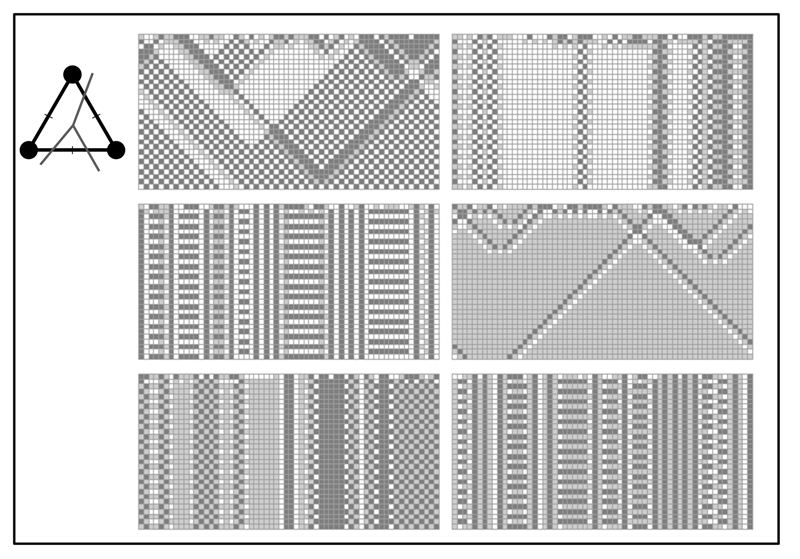}

\includegraphics[scale=0.7]{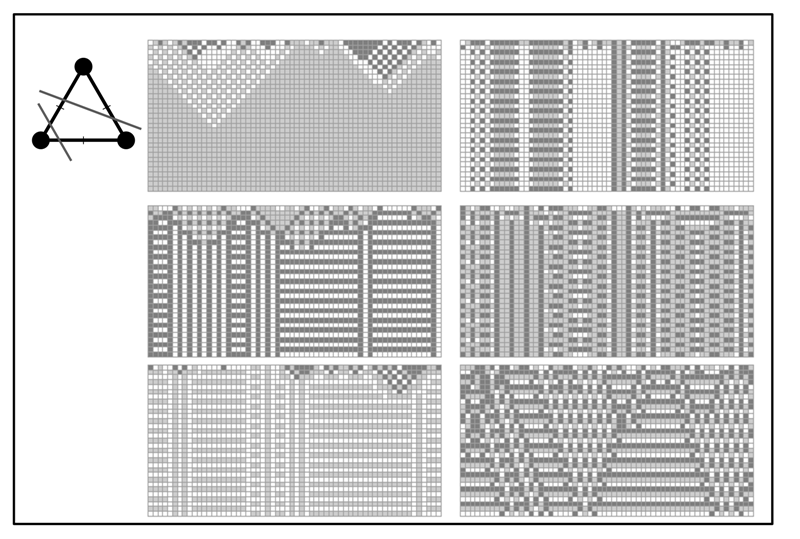}

\includegraphics[scale=0.7]{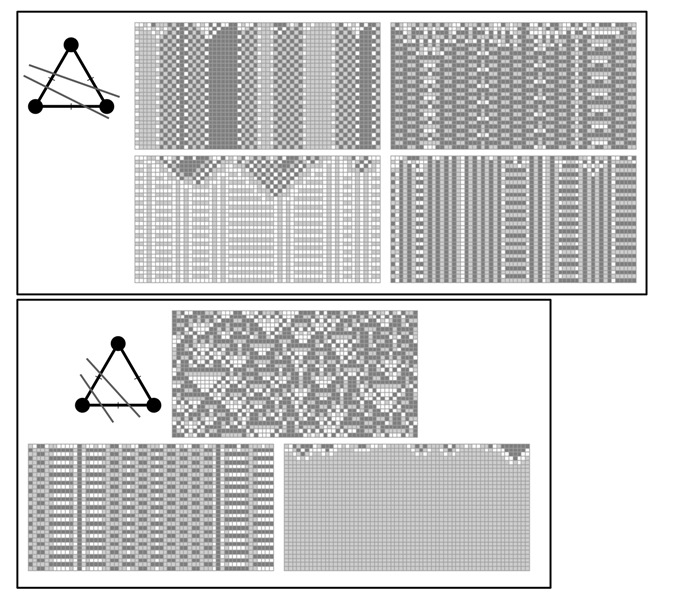}

\includegraphics[scale=0.7]{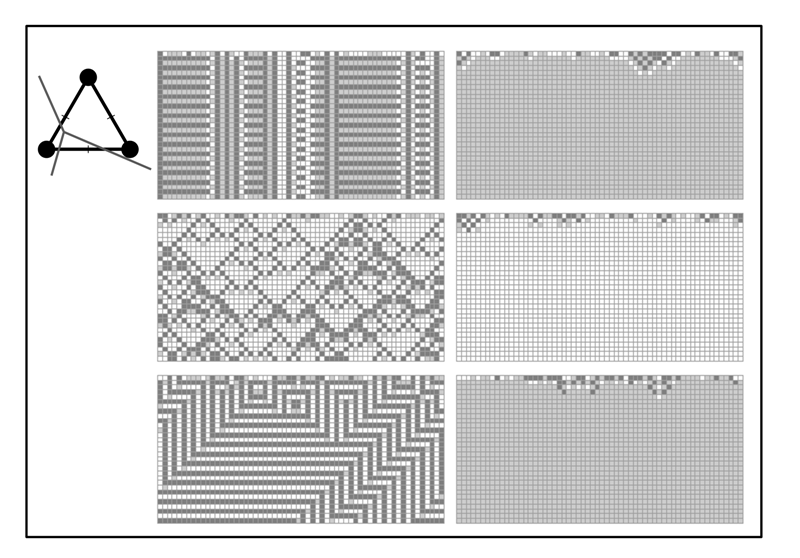}

\ \\

\bibliographystyle{plain} \bibliography{rew}

\end{document}